\documentclass[twocolumn,superscriptaddress,floatfix,preprintnumbers,amssymb ,amsmath]{revtex4}
\usepackage{graphicx}
\usepackage{dcolumn}
\usepackage{bm}
\usepackage[latin1]{inputenc}
\usepackage[mathscr]{eucal}
\usepackage{epsfig}
\usepackage{float}
\usepackage{pdfpages}

 \usepackage{pgfplots}
 \usepackage{tikz}

\begin{document}

\title{Thermodynamics of information exchange between two coupled quantum dots}

\author{Aki Kutvonen}
\affiliation{COMP Center of Excellence, Department of Applied Physics,
Aalto University School of Science, P.O. Box 11000, FI-00076 Aalto, Espoo, Finland}
\author{Takahiro Sagawa}
\affiliation{Department of Applied Physics, The University of Tokyo, 7-3-1 Hongo, Bunkyo-ku, Tokyo, Japan}
\author{Tapio Ala-Nissila}
\affiliation{COMP Center of Excellence, Department of Applied Physics,
Aalto University School of Science, P.O. Box 11000, FI-00076 Aalto, Espoo, Finland}
\affiliation{Department of Physics, Brown University, Providence RI 02912-1843, U.S.A.}

\date{March 13, 2016}

\begin{abstract}
We propose a setup based on two coupled quantum dots where thermodynamics of a measurement can be quantitatively characterized. The information obtained in the measurement can be utilized by performing feedback in a manner apparently breaking the second law of thermodynamics. In this way the setup can be operated as a Maxwell's Demon where both the measurement and feedback are performed separately by controlling an external parameter. This is analogous to the case of the original Szilard engine. Since the setup contains both the microscopic demon and the engine itself, the operation of the whole measurement-feedback cycle can be explained in detail at the level of single realizations. In addition, we derive integral fluctuation relations for both the bare and coarse-grained entropy productions in the setup.
\end{abstract}

\maketitle

\section{INTRODUCTION}
Due to the development of modern nanotechnology, there has been great interest towards measuring and manipulating the microscopic states of physical systems. This has given rise to the field of stochastic thermodynamics characterized by fluctuating thermodynamic variables \cite{Jarzynski2011,Seifert2012,Bustamante2005,Collin2005}. Recently, the role of information and its utilization in small systems have been considered \cite{Sagawa2012b,Sagawa2013,Parrondo2015,Sagawa2010,Sagawa2012,Horowitz2014,Barato2014}. As a result, various theoretical \cite{Ito2015,Horowitz2013, Mandal2013, Strasberg2013, Averin2011, Barato2013}, as well as experimental \cite{Toyabe2010, Koski2014,Koski2014b,Roldan2014} studies have considered how information can be used to lower entropy in apparent contradiction with the second law of thermodynamics. These setups are commonly classified as Maxwell's demons \cite{Leff2003}.

Maxwell's demon is known as an entity which obtains information by measuring the state of the system followed by utilization of information by performing feedback to the system. In the process, entropy of the system decreases while in order to retain the second law the entropy of the surroundings must increase by at least the same amount. Independent of the possible utilization of the information obtained, it is also interesting to study the thermodynamics of the measurement itself. While the process of measurement is ubiquitous in physics, many open questions still remain in its thermodynamic framework, e.g. how a measurement affects the system and how much energy is dissipated in the process. Furthermore, regarding the realizations of the Maxwell's demon, the role of information and thermodynamics in the feedback phase has been studied in detail \cite{Koski2014b,Sagawa2012,Toyabe2010,Parrondo2015}. However, thermodynamics of the measurement phase with a fully physical setup has not been not studied in detail so far. This is partly due to the fact that a rigorous study of the measurement phase requires knowledge of the microscopic state of the entity performing the measurement during the process. Thus the demon, or the measurement device, and its microscopic details and energetics need to be included in the model.

Recently, in order to study the thermodynamics of the demon itself and the operation of the system-demon compound, so called autonomous Maxwell's demon setups, where the demon and system are both built in, have been proposed \cite{Horowitz2014,Strasberg2013,Barato2014,Mandal2013,Shiraishi2015,Koski2015,Kutvonen2015}. In these setups there is no external control and thus the demon acts continuously on the system. The negative entropy production in these systems can be seen as a result of feedback performed by the demon to the system. Flow of information between the demon and the system have been studied \cite{Horowitz2014,Parrondo2015,Shiraishi2015}. In these setups, the demon and the system evolutions are strongly coupled and the demon is measuring the system and affecting its dynamics continuously. Thus a study of a single controlled measurement and its thermodynamics, where the measuring device affects the system only at the point of measurement is not possible. We note that thermodynamics of measurement and feedback in the operation of the Maxwell's demon has also been studied using the framework of information reservoirs \cite{Barato2014,Mandal2012,Defner2013}. In these studies the system is connected to a stream of bits, which manipulates the energy barriers between the system states. However, a physical model with microscopic details allowing to study the measurement phase, and thus the full operation of the device, is yet to be realized.

In this paper, we study a simple model setup where we measure a quantum dot (QD), labeled as the engine dot, with another QD labeled as the measurement dot. We investigate dissipation and information gain in a single measurement event. The information obtained by the measurement dot can be used to perform feedback to the engine dot in a manner apparently breaking the second law of thermodynamics and without the information escaping the engine-measurement dot compound. Thus in the feedback phase the measurement dot can be made to function as a Maxwell's demon. 

To our knowledge this is the first proposal of an experimental setup where both the measurement and feedback are controlled separately, in a non-autonomous fashion, as in the case of the original Szilard engine and the operation of the full measurement-feedback cycle is explained in detail. The setup contains both the microscopic demon and the engine itself and thus the thermodynamics of the full measurement-feedback cycle can be analyzed. Furthermore, we study thermodynamics of the measurement and feedback at the level of single realisations and we derive integral fluctuation relations for both phases. We note that our setup is experimentally realizable, as is the case for an earlier experiment that verified the fluctuation theorem \cite{Utsumi2010}.

This paper is organized as follows. In Section II we introduce the setup. In Section III we focus on thermodynamics of the measurement phase and in Sec. IV on thermodynamics of the feedback phase. Section V addresses fluctuation relations in the setup. Section VI is the summary of the paper.

\section{SETUP}

\begin{figure}{}
    \includegraphics[width=0.33\textwidth]{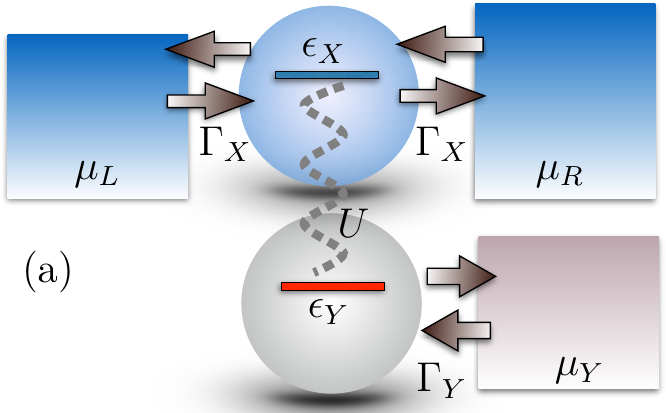}
       \includegraphics[width=0.4\textwidth]{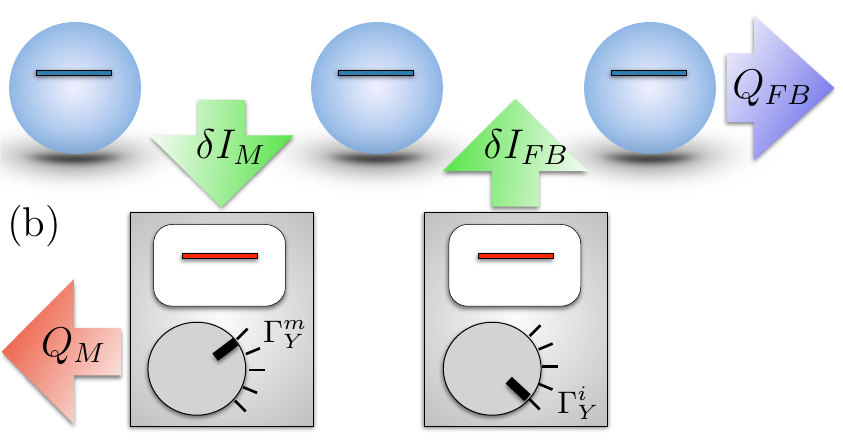}
    \caption{(a): Setup of coupled two-level quantum dots, the engine dot $X$ with energy $\epsilon_{X}$ and the measurement dot $Y$ with energy $\epsilon_{Y}$, coupled to each other so that the state $(x,y)=(1,1)$ has energy $\epsilon_X+\epsilon_Y+U$. State $x$ of the dot $X$ can change due to tunneling in or out from the left (L) and right (R) reservoirs, which are in chemical potentials $\mu_L$ and $\mu_R=\mu_L + \Delta \mu$, respectively, and the coupling strengths are fixed to $\Gamma_X$. Dot $Y$ can change electrons with its reservoir at a chemical potential $\mu_Y$. (b): In the measurement the coupling strength is increased from $\Gamma_Y^{i}$ to $\Gamma_Y^m$ increasing the mutual information between the dots by $\delta I_M$ and dissipating $\beta Q_M \geq \delta I_M$ as heat. In the feedback the coupling strength $\Gamma_Y^m$ is decreased back to its initial value $\Gamma_Y^{i}$ and mutual information $\delta I_{FB}$ is consumed, resulting to negative dissipation $\beta Q_{FB} < 0$, in an apparent violation of the second law. Thus the measurement dot can be made to function as a Maxwell's demon.}
    \label{fig1}
\end{figure}

Our setup consists of two quantum dots, the engine dot $X$ and the measurement dot $Y$ corresponding to the upper and lower dots in Fig. \ref{fig1} (a). The engine dot state $x$, denoting the number of electrons in the dot, may change due to tunneling from left or right reservoir, which are set to chemical potentials $\mu_{L}$ and $\mu_R=\Delta \mu+\mu_L$, respectively. The measurement dot state $y$ may change due to tunneling between the dot and its reservoir at a chemical potential $\mu_Y$. The tunneling rates between the dots and their reservoirs depend linearly on the coupling strengths $\Gamma$, which we denote by $\Gamma_X$ and $\Gamma_Y$ for the engine and the measurement dots, respectively. We operate at low temperatures such that the states of the total combined system are limited to $(x,y) \in \{(0, 0), (0, 1), (1, 0), (1, 1)\}$. The engine and measurement dots have energies $\epsilon_{X}$ and $\epsilon_Y$, respectively, when filled, and zero energy when empty. Furthermore, the dots are capacitively coupled by an interaction energy $U$, so that the energy of the state $(1,1)$ is given by $\epsilon_{X}+\epsilon_Y+U$. 

We perform the measurement by increasing the coupling strength $\Gamma_Y$, from its initial value $ \Gamma_Y^{i}$ to its measurement value $\Gamma_Y^{m}$, increasing the relaxation rate to energetically favorable total energy states. As the probability of energetically favorable states is increased in the process, the correlation between states $x$ and $y$, quantified by the mutual information $\langle I(x,y) \rangle = \sum_{x,y}p(x,y) \ln [p(x,y) /(p_S(x) p_Y(y))]$, increases \cite{Parrondo2015}. Here $p_{X}(x)=\sum_{x}p(x,y)$ and $p_{Y}(y)=\sum_{x}p(x,y)$ are the marginal distributions of the engine and the measurement device, respectively. After the measurement phase, the measurement dot can be made to work as a Maxwell's demon by decreasing the coupling strength $\Gamma_Y$ from its measurement value $\Gamma_Y^m$ back to the initial value $\Gamma_Y^{i}$. In this feedback phase mutual information $I$ is utilized resulting to negative engine entropy production. Both the measurement and feedback are schematically illustrated in Fig. \ref{fig1} (b). We note that the external control is done in a deterministic fashion and thus there is no exchange of information or entropy associated to the control. Therefore, the external control does not operate as a "super-demon" in the present setup.

Continuing to the energetics of the setup, dissipation in a tunneling event $(x_k,y_k) \to (x_l,y_l)$, over the left ($\upsilon=L$), the right ($\upsilon=R$), or the measurement ($\upsilon=Y$) junction, is given by the energy released $q_\upsilon^{k \to l}=\Delta x (\epsilon_X-\mu_{L/R})+\Delta y (\epsilon_Y-\mu_Y)+(x_k y_k-x_l y_l) U$, where $\Delta x = x_k-x_l$, $\Delta y = y_k-y_l$ and $\mu_{L/R}$ should be selected for tunneling over the left/right junction. Furthermore, we denote the reversed tunneling direction $k \leftarrow l$ so that $q_\upsilon^{k \to l}=-q_\upsilon^{k \leftarrow l}=-q_\upsilon^{l \to k}$. The tunneling rates are given by
\begin{equation}
\omega^{k \rightleftarrows l}= \Gamma \frac{1}{1+e^{-\beta q^{k \rightleftarrows l}}},
\label{eq:rates}
\end{equation}
where $\beta=1/(k_B T)$. The rates $\omega$ are local detailed balance (LDB) connected so that $q^{k \to l}=\beta^{-1} \ln [\omega^{k \to l}/\omega^{l \to k}]$.   

The probability distribution of state $(x_k,y_k)$, follows the master equation
\begin{equation}
\frac{d p_k}{dt}=\sum_{l}[ p_l \omega^{l \to k}  -p_k \omega^{k \to l}].
\label{eq:Meq}
\end{equation}
By conveniently setting the chemical potentials to $\epsilon_Y=\mu_Y+U/2$ and $\epsilon_X=(\mu_L+\mu_R)/2-U/2$, the dissipation in tunneling events become degenerate:  $q_{L/R}^{s \to a} \equiv q_{L/R}^{(0,0)\to(1,0)}=q_{R/L}^{(1,1)\to(0,1)}=1/2[\pm \mu_L \mp \mu_R +U]$ and $q_{Y}^{s \to a} \equiv q_Y^{(1,1)\to(1,0)}=q_Y^{(0,0)\to(0,1)}=U/2$. Thus the rates  $\omega$ (Eq. (1)) then also become degenerate: ${}^Y \omega_{c}^{s \rightleftarrows a}= \omega_{c}^{(1,1) \rightleftarrows (1,0)}=\omega_{c}^{(0,0) \rightleftarrows (0,1)}$ and ${}^X \omega^{s \rightleftarrows a} \equiv {}^L \omega^{(0,0) \rightleftarrows (1,0)}+{}^R \omega^{(0,0) \rightleftarrows (1,0)}={}^L \omega^{(1,1) \rightleftarrows (0,1)}+{}^R \omega^{(1,1) \rightleftarrows (0,1)}$. Therefore the antisymmetric states $a$, $(0,1)$ and $(1,0)$, and correspondingly the symmetric states $s$, $(0,0)$ and $(1,1)$, become equally probable to occupy. Furthermore, we denote the effective rate at which the states $s$ and $a$ change by $\omega^{s \rightleftarrows a}={}^Y \omega^{s \rightleftarrows a}+{}^X \omega^{s \rightleftarrows a}$. By the capital letter $P$ we denote the probability distribution of states $s$ and $a$, where $P_s=2p(0,0)=2p(1,1)=\omega^{a \to s}/Z$ and $P_a=2p(0,1)=2p(1,0)=\omega^{s \to a}/Z$, and where $Z=\omega^{s \to a}+\omega^{a \to s}$. Thus the total system effectively becomes a two-level system with the energetically favored antisymmetric state $a$ and the symmetric state $s$.

\section{MEASUREMENT}

\begin{figure}{}
    \includegraphics[width=0.3\textwidth]{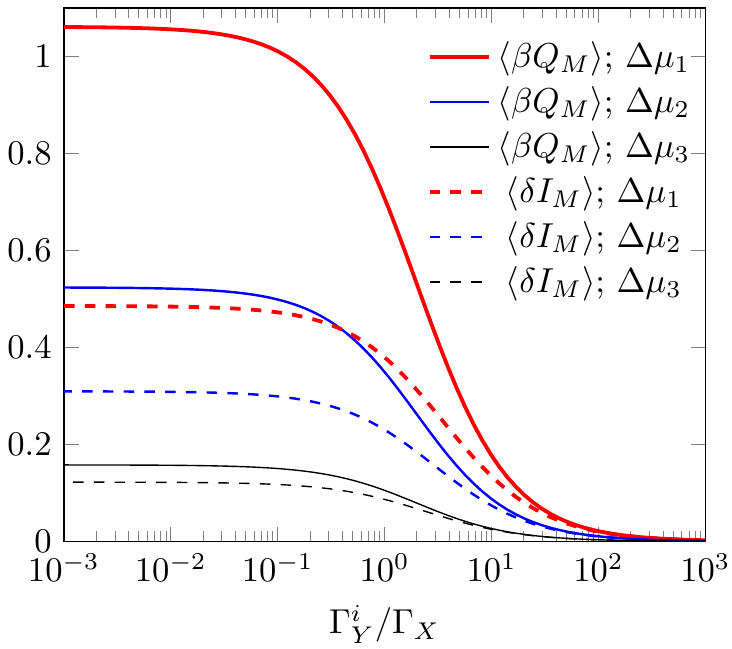}
    \caption{Average dissipation $\langle \beta Q_M \rangle$ and the change in the mutual information $\langle \delta I_M \rangle$ in the measurement as a function of the initial coupling strength $\Gamma_Y^{i}$ with three different biases, $\Delta \mu_1=0.95U$, $\Delta \mu_2=0.75U$ and $\Delta \mu_3=0.5U$ assuming that the post measurement state is thermalized. The bias $\Delta \mu_2$  and the temperature used $\beta U = 10$ correspond to the optimal values for negative entropy production in the feedback phase, extracted from data shown in Fig. \ref{fig4} (a). $\langle \delta I_M \rangle >0$, showing that correlations are built. Furthermore, entropy production $\langle S_M \rangle=\beta \langle Q_M-\delta I_M \rangle \geq 0$ is positive. For coupling strengths higher than $\Gamma_Y \simeq10^2$ and lower than $\Gamma_Y \simeq 10^{-2}$ neither $\langle \delta I_M \rangle$ or $\langle \beta Q_M \rangle$ change, which signals thermalization of the measurement dot and the engine, respectively. In the main text we consider the case of a small $\Gamma_Y^{i}$ so that the whole measurement-feedback process can be made cyclic.}
     \label{fig2}
\end{figure}

Initially the measurement dot is prepared to a state at a weak coupling strength $\Gamma_Y^{i} \ll \Gamma_X$ which leads to a steady state distribution $P^{i}$. We assume that the coupling strength $\Gamma_Y^{i}$ is weak enough so that the engine dot is thermalized. By this we mean that transition timescales in the engine dot are much faster than those of the measurement dot. Thus the system dot relaxes to a steady state non-equilibrium distribution and not to a thermal equilibrium state. Numerically the thermalized limit is obtained roughly at $\Gamma_Y^{i} \approx 10^{-2} \Gamma_X$, as shown in Fig \ref{fig2}. In this case the steady state distribution is given by
\begin{equation}
P^{i}_{s/a} = [1+e^{\sigma_{FB}^{s \rightleftarrows a}}]^{-1},
\label{eq:Pfeedback}
\end{equation}
where $\sigma_{FB}^{s \rightleftarrows a}=\ln[{}^X \omega_{fb}^{s \rightleftarrows a} /{}^X \omega_{fb}^{a \rightleftarrows  s}]$ is the coarse-grained entropy. Because the engine state $x$ may change through the left or right junctions, the rates ${}^X \omega$ are not detailed balance connected and the coarse-grained entropy $\sigma_{FB}$ does not equal the bare medium entropy $\beta q_{L/R}$.

 After rapid increase of the coupling strength from $\Gamma_Y^{i}$ to $\Gamma_Y^{m} \gg\Gamma_Y^{i}$ the total combined system is allowed to relax to a new steady state $P^m$. The new value of the coupling strength, $\Gamma_Y^{m}$, is set high enough ($\Gamma_Y^{m} / \Gamma_X > 10^2$) so that the measurement dot thermalises. Furthermore, the measurement time $\tau_{m}$ from the beginning of the change of $\Gamma_Y$ up to the time of relaxation to $P^m$ is set short enough ($ \ll \Gamma^{-1}_X$) such that during the measurement the state of the engine does not change. If there were tunneling events in the engine dot, they would cause additional entropy production through dissipated heat. However, by setting the coupling strength and measurement time so that $(\Gamma^{-1}_X)^{-1} \gg \tau_m \gg  (\Gamma_Y^{m})^{-1}$, this possibility can be neglected.
 
The ratio of transition probabilities $T_m$ between states $a$ and $s$ during the measurement is given by $T_m^{s \to a}/T_m^{a \to s}={}^Y \omega_m^{s \to a}/{}^Y \omega_m^{a \to s}=e^{\beta q_Y^{s \to a}}$, where ${}^Y \omega_m^{s \rightleftarrows a}= \omega_m^{(1,1) \rightleftarrows (1,0)}= \omega_m^{(0,0) \rightleftarrows (0,1)}$. The thermalized steady state probability distribution is given by
\begin{equation}
P^{m}_{s/a} = [1+e^{\beta q_Y^{s \rightleftarrows a}}]^{-1}.
\label{eq:Pmeas}
\end{equation}

The non-equilibrium free energy of the engine-measurement device compound is given by $F_{X+Y}=\langle E_{X+Y} \rangle+\beta^{-1}\sum _{x,y} p(x,y) \ln [p(x,y)]$, where $\langle E_{X+Y} \rangle=\sum _{x,y} p(x,y) E_{X+Y}$ is the total energy and the latter term is the Shannon entropy \cite{Esposito2011}. By splitting the total energy into the engine contribution $\langle E_X \rangle=\sum_x p_{X}(x)E_X(x)$, the measurement device contribution $\langle E_Y \rangle= \sum_y p_{Y}(y)E_Y(y)$ and the interaction part $\langle E_I \rangle = \sum _{x,y} p(x,y) E_{k}(x,y)$, the free energy can be written as 
\begin{equation}
F_{X+Y}=
F_X+F_Y+\langle E_I\rangle +\beta^{-1} \langle I \rangle,
\end{equation}
where $F_X=\langle E_X \rangle+\beta^{-1}\sum _{x} p_X(x) \ln [p_X(x)]$ and $F_Y=\langle E_Y \rangle+\beta^{-1}\sum _{y} p_Y(y) \ln [p_Y(y)]$. The total entropy production in the manipulation of the total combined system is given by $\langle S_M \rangle =\beta [W -\delta F_{X+Y}]$, where $W$ is the external work done in performing the measurement, i.e. in changing the coupling strength $\Gamma_Y$. By assuming that no dissipation occurs outside the compound system in switching the coupling strength $\Gamma_Y$, the measurement work is given by the sum of energy increase in the total system and the dissipated heat. The probability of observing, for example, the state $p_X(1)$ is given by $p(1,0)+p(1,1)=1/2[P_a+P_s]=1/2$ both before and after the measurement. Similarly $p_X(x)=p_Y(y)=1/2$ for all $x\in{0,1}$ and $y\in{0,1}$. That is to say, the marginal distributions do not change in the process and therefore in both the measurement and feedback the terms $E_X$, $E_Y$, $F_X$, $F_Y$ do not change. Therefore the entropy production in the measurement is given by
\begin{equation}
\langle S_M \rangle = \langle \beta Q_M \rangle- \langle \delta I_M \rangle \geq 0,
\label{eq:MeasurementEntropy}
\end{equation}
where we used the fact the measurement work is given by $W=\delta E_I+Q_M$, where $Q_M$ is the heat dissipated in the measurement. Since we assumed that there are no tunneling events through the engine dot during the measurement, the dissipated heat equals the decrease in the internal energy, $Q_M=-\delta E_I$ and therefore the measurement work is zero. This is consistent with the famous Landauer principle which states that work is needed for the erasure of memory, but the measurement itself can be done without external work \cite{Landauer1961}.

The average heat dissipated in the measurement is given by
\begin{equation}
\begin{aligned}
\langle Q_M \rangle &=P^{i}_sT_{m}^{s \to a}q_Y^{s \to a}+P^{i}_aT_{m}^{a \to s}q_Y^{a \to s}=\frac{U}{2}\Delta P_{m},
\label{eq:Qm}
\end{aligned}
\end{equation}
where $\Delta P_{m}=P^{m}_a-P^{i}_a=T_{m}^{s \to a}P^{i}_s-T_{m}^{a \to s}P^{i}_a$ is the change in the probability of state $a$ and $q_Y^{s \rightleftarrows a}=\pm U/2$. Because the probability of occupying the energetically favourable state $a$ increases in the measurement, $\langle Q_M \rangle \geq 0$ as can be also seen from Fig. \ref{fig2}.

The mutual information change in the measurement is given by $\delta I_M^{k \to l}=\ln  [p^{m}(x_l,y_l)p^{i}_Y(y_k)]/[p^{i}(x_k,y_k)p^{m}_Y(y_l)]\}$, which due to the symmetry of the marginal distributions, $p^{i}_Y(y)=p^{m}_Y(y)=1/2$, reduces to $\delta I_M^{k \to l}=\ln \frac{P^{m}_l}{P^{i}_k}$, where $k,l\in\{a,s\}$. The average value of $\delta I_M$ is given by
\begin{equation}
\begin{aligned}
&\langle \delta I_M \rangle 
=P^{i}_s\ln \frac{P^{m}_s}{P^{i}_s}+P^{i}_a\ln \frac{P^{m}_a}{P^{i}_a}+\langle \beta Q_M \rangle
\label{eq:avIm}
\end{aligned}
\end{equation}
as shown in detail in Appendix A.

\section{FEEDBACK}

Next we study the thermodynamics of the feedback phase, by decreasing the coupling strength from $\Gamma_Y^m$ back to the initial value $\Gamma_Y^{i}$ and making the process cyclic. During the feedback mutual information $I$ is utilized to produce negative entropy in the form of cooling. Analogous to the measurement phase, the feedback time $\tau_{fb} \ll (\Gamma_Y^{f})^{-1}$ is set such that the measurement dot state does not change during the feedback phase.

Similar to the measurement phase, the symmetry $p_X(x)=p_Y(y)=1/2$ is preserved even if the state $x$ changes and thus the entropy production is given by
\begin{equation}
\langle S_{FB} \rangle =\langle \beta Q_{FB} \rangle - \langle \delta I_{FB} \rangle \geq 0,
\label{eq:FeedbackEntropy}
\end{equation}
where $Q_{FB}$ is the heat dissipated from the engine during the feedback phase. We note that from the global perspective, the energy source which fuels the change in the chemical potential is drained in the feedback phase. This energy source performs work to the system during the feedback. However, when observing the engine only, the second law of thermodynamics is apparently violated. 

\subsection{Heat from relaxation}

\begin{figure}{}
    \includegraphics[width=0.24\textwidth]{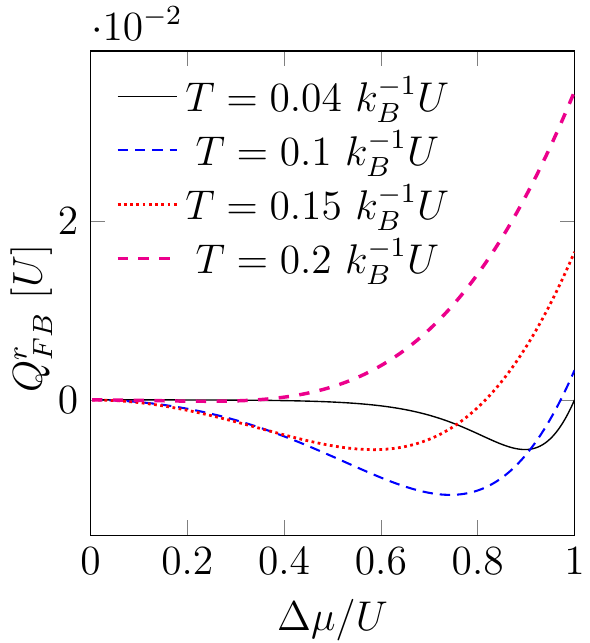}
       \includegraphics[width=0.23\textwidth]{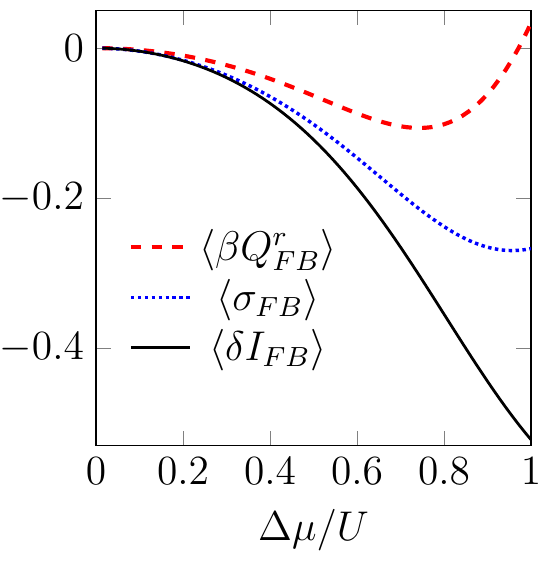}
    \caption{(a): Average relaxation heat  in feedback $\langle Q_{FB}^r \rangle$ as a function of bias to the interaction ratio $\Delta \mu / U$ at different operation temperatures $T$. The minimum of $\langle Q_{FB}^r \rangle$ is obtained roughly at $\Delta \mu=0.75 U$ and $T=0.1 ~k^{-1}_B U$. (b):  Relaxation heat entropy $\langle \beta Q_{FB}^r \rangle$, coarse-grained entropy $\langle \sigma_{FB} \rangle$ and mutual information $\langle \delta I_{FB} \rangle$ production in the feedback at $T=0.1 ~k^{-1}_B U$. Since $\langle \beta Q_{FB}^r \rangle \geq  \langle \sigma_{FB} \rangle \geq \langle \delta I_{FB} \rangle$, the negative entropy production is bounded by the change in the mutual information $-\beta^{-1} \langle \delta I_{FB} \rangle$. In both (a) and (b) we assumed that the post-feedback state $P(\tau_{fb})$ is the thermalized initial state $P^i$ of Eq. \eqref{eq:Pfeedback}. Results are obtained by numerically solving the master equation of Eq. \eqref{eq:Meq}.}
    \label{fig3}
\end{figure}

We look at the total dissipated heat in the feedback as a sum of heat caused by the relaxation of $P$ from the pre-feedback state $P^m$ to the post-feedback state $P(\tau_{fb})$, given by $Q_{FB}^r$, and heat caused by a continuous current of electrons through the dot, $Q_{FB}^c$, i.e. $Q_{FB}=Q_{FB}^r+Q_{FB}^c$. The relaxation heat is given by 
\begin{equation}
\begin{aligned}
\langle Q_{FB}^r \rangle &=[q_L^{s \to a}P_{L|s \to a}+q_R^{s \to a}P_{R|s \to a}]P^{m}_sT_{fb}^{s \to a} \\
&+[q_L^{a \to s}P_{L|a \to s}+q_R^{a \to s}P_{R|a \to s}]P^{m}_aT_{fb}^{a \to s}, \\
\label{eq:Qfb}
\end{aligned}
\end{equation}
where $P_{L/R | s\rightleftarrows a}={}^{L/R} \omega^{(0,0) \rightleftarrows (1,0)}/{}^{X}\omega^{s \rightleftarrows a}={}^{L/R} \omega^{(1,1) \rightleftarrows (0,1)}/{}^{X}\omega_{s \rightleftarrows a}$ is the conditional probability to tunnel over the left/right junction given the state change $s \rightleftarrows a$. The ratio of transition probabilities is given by the ratio of the transition rates $T_{fb}^{s \to a}/T_{fb}^{a \to s}={}^X \omega_{fb}^{s \to a}/{}^X \omega_{fb}^{a \to s}=e^{\sigma_{FB}^{s \to a}}$. However, the absolute values of the transition probabilities depend on the feedback time  $\tau_{fb}$. The shorter the $\tau_{fb}$, the less time the engine has to re-equilibrate making the transition probabilities $T_{fb}$ smaller. In the limit of $\tau_{fb}=0$ the transition probabilities $T_{fb}$ are zero. In the other limit when $\tau_{fb} \approx \Gamma_X^{-1}$ is set long enough, the engine relaxes to the initial thermalized pre-measurement probability distribution $P^{i}_{s/a}$ (Eq. \eqref{eq:Pfeedback}). In this case the transition probabilities are given by $T_{fb}^{s \rightleftarrows a}=[1+e^{-\sigma_{FB}^{s \rightleftarrows a}}]^{-1}$. In Fig. \ref{fig3} (a) we plot the relaxation heat $\langle Q_{FB}^r \rangle$ as a function of temperature and bias $\Delta \mu$ in the thermalized limit.

\subsection{Heat from electron current}

Within a relatively large parameter range we are able to find a regime where the relaxation heat $\langle Q_{FB}^r \rangle$ is negative. This cooling effect competes against heating caused by continuous energy dissipation due to the migration of the electrons from the higher chemical potential $\mu_R$ to the lower $\mu_L$, which heats the engine by $\Delta U$ for each electron transferred through the engine dot. The crucial question regarding the operation as a Maxwell's demon device is whether the total dissipated heat $\langle Q_{FB} \rangle$ is negative or not. The total heat dissipation rate 
\begin{equation}
\begin{aligned}
\langle \dot Q_{FB} [P(t)]\rangle
=P(t)_s \dot Q_{s \to a}+P(t)_a\dot Q_{a \to s}, \\
\end{aligned}
\end{equation}
in the feedback phase depends on the instantaneous probability distribution of the states $P(t)$. Here $\dot Q_{s \to a}=[q_L^{s \to a}P_{L|s \to a}+q_R^{s \to a}P_{R|s \to a}]{}^X \omega_{s \to a}$ and $\dot Q_{a \to s}=[q_L^{a \to s}P_{L|a \to s}+q_R^{a \to s}P_{R|a \to s}]{}^X \omega_{a \to s}$.

\begin{figure}{}
    \includegraphics[width=0.35\textwidth]{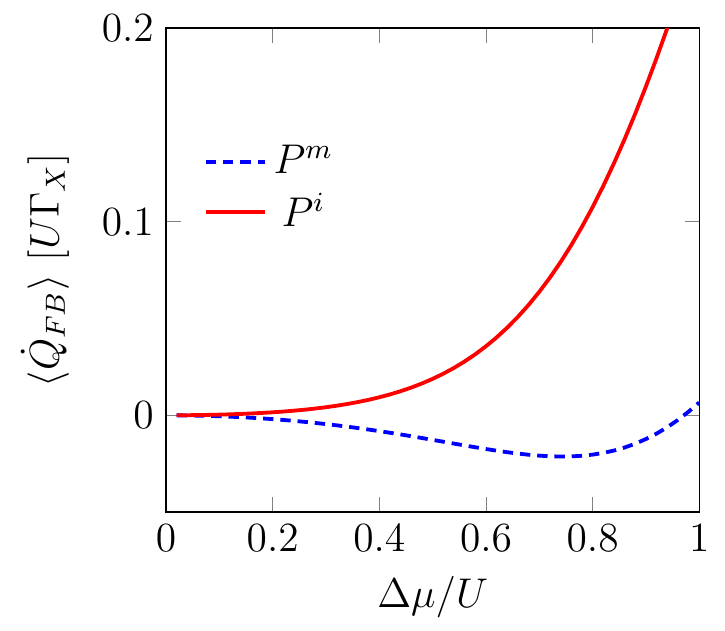}
    \caption{Heat production rate $\langle \dot Q_{FB} \rangle$ in feedback in two extreme cases as a function of bias to interaction energy ratio with interaction energy $U=10~k_{\rm B} T$. The maximum cooling rate $\langle \dot Q_{FB}[P^m] \rangle$ is obtained immediately after the switching of the coupling parameter from $\Gamma_Y^m$ back to its initial value $\Gamma_Y^i$. When the engine relaxes after time $t_r$ to the thermalized steady state $P(t_r)=P^i$ the cooling rate $\langle \dot Q_{FB}[P^m] \rangle$ becomes positive. The total heat dissipation is bounded by $\langle \dot Q_{FB}[P^m] \rangle \tau_{fb} \leq \langle Q_{FB} \rangle \leq \langle \dot Q_{FB}[P^m] \rangle \tau_{fb}$. By making the feedback time short enough, one can ensure that $\langle \dot Q_{FB}[P(\tau_{fb})] \rangle$ is negative and thus the total dissipation of heat in the feedback $\langle  Q_{FB} \rangle$ is negative.}
    \label{fig4}
\end{figure}

Immediately after the change of the coupling strength $\Gamma_Y^m$ back to its initial value $\Gamma_Y^i$, the probability distribution of electrons on the engine dot is given by $P^m$ and thus the heat production rate is given by $\langle \dot Q_{FB}[P^m] \rangle$. This heat production rate is essentially the heat production of Eq. \eqref{eq:Qfb} multiplied by a constant, and can thus be negative as shown in Figs. \ref{fig3} and \ref{fig4}. Cooling is possible because of the higher probability of the antisymmetric state $a$, $P^m_a$, in the measurement phase. As the engine relaxes towards the new steady state, the probability of state $a$ increases and the probability distribution $P(t)$ and the heat production rate $\langle \dot Q_{FB}(t) \rangle$ changes in time. After a long enough time $t=t_r$, the engine has relaxed to $P(t_r)=P^i$ and the new steady state heat production rate $\langle \dot Q_{FB}[P^i] \rangle$ is positive. Here we denoted $t_r=\Gamma_X^{-1}$ as the relaxation time of the engine. Thus, if the feedback time is set too long, $\tau_{fb} \gg t_r$, the steady state heating caused by the current will overrun the initial cooling. In this case the total heat becomes positive, that is to say $\langle Q_{FB} \rangle=\langle Q_{FB}^r \rangle +\langle Q_{FB}^c \rangle >0$.

However, by setting the feedback time shorter than the relaxation time of the engine, $\tau_{fb} \ll t_r$, one can ensure that the contribution to the total dissipated heat from the cooling regime where $\langle \dot Q_{FB} [P(t)]\rangle <0$ dominates over the heating regime where $\langle \dot Q_{FB} [P(t)]\rangle >0$ and the setup operates as a Maxwell's demon device. By making the feedback extremely short, one can make sure that during the whole feedback the engine operates at the cooling regime, such that $\langle \dot Q_{FB} [P(\tau_{fb})]\rangle <0$. In general the cooling and heating regimes depend on the parameters, but the two extremes for the total heat production are given by the maximum heating $\langle \dot Q_{FB} [P^i]\rangle$ and maximum cooling $\langle \dot Q_{FB} [P^m]\rangle$, shown for interaction strength $U=10 ~k_{\rm B} T$ in Fig. \ref{fig4}. 

\subsection{Cyclic operation and mutual information}

In the cyclic operation of the device, the feedback time also has an effect on the measurement phase. If $\tau_{fb}$ is short, the probability distribution $P(\tau_{fb})$ after the first feedback phase is still close to $P^m$ and not given by the initial probability distribution $P^i$. Thus in the next cycles the initial probability distribution $P^i$ should be replaced by the distribution $P(\tau_{fb})$ in Eqs. \eqref{eq:avIm} and \eqref{eq:Qm}. Compared to the measurement phase in the first cycle, in the following cycles there will be less relaxation and thus less dissipation and change in the mutual information.

The change of mutual information in the feedback is given by $\delta I_{FB}^{s \rightleftarrows a}=\ln [P(\tau_{fb})_{a/s}/P^{m}_{s/a}]$. The absolute value of the change of mutual information depends on the change of $P$ during the feedback, which depends on the feedback time. The longer the feedback time, the more $P$ changes resulting to a larger change in the mutual information. In the limit of $\tau_{fb} \to 0$, $P(\tau_{fb}) \to P^m$ and thus $\langle \delta I_{FB} \rangle \to 0$. In the thermalized limit, when the feedback time is long enough ($\tau_{fb} \simeq t_r$) to allow the engine to relax to the thermalized state $P^i$, the average of $\delta I_{FB}$ is given by
\begin{equation}
\begin{aligned}
&\langle \delta I_{FB} \rangle
=P^{m}_s\ln \frac{P(\tau_{fb})_s}{P^{m}_s}+P^{m}_a\ln \frac{P(\tau_{fb})_a}{P^{m}_a}+\langle \sigma_{FB} \rangle
\label{eq:avIfb}
\end{aligned}
\end{equation}
as shown in Appendix B in detail.

\section{FLUCTUATION THEOREMS}

By a straightforward use of the definitions, shown in Appendix C in detail, both the relaxation heat entropy $\beta Q_{FB}$ and the coarse-grained entropy $\sigma_{FB}$ together and separately satisfy the integral fluctuation relations:
\begin{equation}
\langle e^{-\beta Q_{FB}^r+\delta I_{FB}} \rangle 
=\langle e^{-\sigma_{FB}+\delta I_{FB}} \rangle
=\langle e^{-\beta Q_{FB}^r+\sigma_{FB}} \rangle
=1.
\label{eq:IFTFeedb}
\end{equation}
By using the Jensen inequality to the equations above we obtain $\langle \beta Q^r_{FB} \rangle \geq  \langle \sigma_{FB} \rangle \geq \langle \delta I_{FB} \rangle$. Furthermore, we know that $Q_{FB} \geq Q_{FB}^r$ and thus the maximum amount of negative entropy production is given by the decrease in the mutual information, and the coarse-grained entropy production always underestimates the entropy production through relaxation heat. Since the negative heat $Q_{FB}$ could be used for source of work extraction, as is the case for extraction of mechanical work out of isothermal expansion in the Szilard engine, the possible extracted work is bounded by the change in mutual information $-\delta I_{FB}/\beta$. In Figure \ref{fig3} we plot the entropy productions $\langle \beta Q^r_{FB} \rangle$ and $\langle \sigma_{FB} \rangle$, and the change in the mutual information $\langle \delta I_{FB} \rangle$ assuming that the post feedback state is the initial thermalized state, that is to say $P(\tau_{fb})=P^i$.

In addition, as shown in Appendix D in detail, the measurement entropy obeys the integral fluctuation theorem:
\begin{equation}
\begin{aligned}
&\langle e^{-\beta Q_M+\delta I_M} \rangle =1,
\label{eq:IFTMeas}
\end{aligned}
\end{equation}
which by using Jensen's inequality is consistent with Eq. \eqref{eq:MeasurementEntropy}. 

\section{SUMMARY}

In summary, we have studied thermodynamics of measurement, in particular dissipation and information gain by measuring a biased quantum dot with another quantum dot. The thermodynamic cost of obtaining information is dissipation in the measurement device so that the second law of thermodynamics is satisfied for the whole engine-measurement device compound. We have shown that the same setup can work as a Maxwell's demon device which uses mutual information obtained in the measurement to produce negative entropy, which is bounded by the decrease in mutual information. The measurement and feedback are performed separately as in the case of the original Szilard engine, and the setup contains both the engine and the demon. Therefore the model contains the full measurement-feedback cycle in a transparent way.  In addition we have shown analytically that both the bare and coarse-grained entropy productions satisfy integral fluctuation relations separately in measurement and feedback. Physical realizations of our model requires a quantum dot setup where the coupling strengths can be manipulated. This can be done in QDs defined and controlled by electrostatic gates, where transparency of tunnel barriers can be adjusted using electric fields \cite{Wiel2002}. Interesting future research directions include studying the same setup in the quantum regime.

\section*{ACKNOWLEDGMENTS}

This research has been supported by the Academy of Finland through its Centres of Excellence Program (project nos. 251748 and 284621), JSPS KAKENHI Grant No. 25800217 and No. 22340114, and by KAKENHI No. 25103003 "Fluctuation \& Structure". We wish to thank Jukka Pekola, Eiki Lyoda, Russell Lake and Samu Suomela for useful discussions.

\section*{APPENDIX A: DERIVATION OF EQ. \eqref{eq:avIm}}

The average value of mutual information obtained in the measurement is given by
\begin{equation}
\begin{aligned}
&\langle \delta I_M \rangle =P^{i}_s [T_{m}^{s \to a}\delta I_M^{s \to a}\\
&+T_{m}^{s \to s}\delta I_M^{s \to s}]+P^{i}_a[T_{m}^{a \to s}\delta I_M^{a \to s}+T_{m}^{a \to a}\delta I_M^{a \to a}],\\
\end{aligned}
\end{equation}
where the terms $\delta I_M^{s \to s}$ and $\delta I_M^{a \to a}$ exists because the mutual information changes in the measurement even if the state does not change, since the probability distribution $P^{i} \to P^m$ changes. By using the fact that $T_{m}^{s/a \to a/s}=1-T_{m}^{s/a \to s/a}$ and $\delta I_M^{s \to a}=\ln P^{m}_a /P^{i}_s$, we obtain
\begin{equation}
\begin{aligned}
&\langle \delta I_M \rangle =P^{i}_s\{T_{m}^{s \to a}\ln \frac{P^{m}_a}{P^{m}_s}+\ln \frac{P^{m}_s}{P^{i}_s} \}\\
&+P^{i}_a\{T_{m}^{a \to s}\ln \frac{P^{m}_s}{P^{m}_a}+\ln \frac{P^{m}_a}{P^{i}_a} \} \\
&=\langle \beta Q_M \rangle + P^{i}_s\ln \frac{P^{m}_s}{P^{i}_s}+P^{i}_a\ln \frac{P^{m}_a}{P^{i}_a}, \\
\end{aligned}
\end{equation}
where we used the assumption of the relaxed post measurement state $P^m$ (Eq. (4)) and the definition of $\langle Q_M \rangle$ (Eq. (7)).

\section*{APPENDIX B: DERIVATION OF EQ. \eqref{eq:avIfb}}

The average mutual information change in the feedback is given by
\begin{equation}
\begin{aligned}
&\langle \delta I_{FB} \rangle=P^{m}_s[T_{fb}^{s \to a}\delta I_{FB}^{s \to a}+T_{fb}^{s \to s}\delta I_{FB}^{s \to s}]\\
&+P^{m}_a[T_{fb}^{a \to s}\delta I_{FB}^{a \to s}+T_{fb}^{a \to a}\delta I_{FB}^{a \to a}]. \\
\end{aligned}
\end{equation}
By using the fact that $T_{fb}^{s/a \to a/s}=1-T_{fb}^{s/a \to s/a}$ and  $\delta I_{FB}^{s \to a}=\ln P(\tau_{fb})_a / P^{m}_s$ we obtain
\begin{equation}
\begin{aligned}
&\langle \delta I_{FB} \rangle=P^{m}_s\{T_{fb}^{s \to a}\ln \frac{P(\tau_{fb})_a}{P^{m}_s}+\ln \frac{P(\tau_{fb})_s}{P^{m}_s} \}\\
&+P^{m}_a\{T_{fb}^{a \to s}\ln \frac{P(\tau_{fb})_s}{P^{m}_a}+\ln \frac{P(\tau_{fb})_a}{P^{m}_a} \}. \\
\end{aligned}
\end{equation}
As discussed in the main text, the transition probabilities $T_{fb}^{a \rightleftarrows s}$ are bounded by $0 \leq T_{fb}^{a \rightleftarrows s}\leq [1+e^{-\sigma_{FB}^{a \rightleftarrows s}}]^{-1}$, where the lower limit is obtained in the case of short feedback time and the upper one in the case of long feedback ($P(\tau_{fb})=P^i$). In the relaxed limit we obtain
\begin{equation}
\begin{aligned}
&\langle \delta I_{FB} \rangle=\langle \sigma_{FB} \rangle + P^{m}_s\ln \frac{P^{i}_s}{P^{m}_s}+P^{m}_a\ln \frac{P^{i}_a}{P^{m}_a},\\
\end{aligned}
\end{equation}
where we used the definition $P^{i} ($Eq. (3)). 

\section*{APPENDIX C: DERIVATION OF EQ. \eqref{eq:IFTFeedb}}
\begin{widetext}
The integral fluctuation theorem for the feedback entropy is given by
\begin{equation}
\begin{aligned}
&\langle e^{-\beta Q_{FB}^r+\delta I_{FB}} \rangle  =P^{m}_s\{T_{fb}^{s \to a}[P_{L|s \to a}e^{-\beta q_L^{s \to a}}+P_{R|s \to a}e^{-\beta q_R^{s \to a}}]e^{\delta I_{FB}^{s \to a}}+T_{fb}^{s \to s}e^{\delta I_{FB}^{s \to s}}\} \\
&+P^{m}_a\{T_{fb}^{a \to s}[P_{L|a \to s}e^{-\beta q_L^{a \to s}}+P_{R|a \to s}e^{-\beta q_R^{a \to s}}]e^{\delta I_{FB}^{a \to s}}+T_{fb}^{a \to a}e^{\delta I_{FB}^{a \to a}}\}.
\label{AD1}
\end{aligned}
\end{equation}
By using the identity
\begin{equation}
\begin{aligned}
 P_{L|s \rightleftarrows a}e^{-\beta q_L^{s \rightleftarrows a}}+P_{R|s \rightleftarrows a}e^{-\beta q_R^{s \rightleftarrows a}}
&=\frac{{}^L \omega^{s \rightleftarrows a}}{{}^X \omega^{s \rightleftarrows a}}e^{-\beta q_L^{s \rightleftarrows a}}+\frac{{}^R \omega^{s \rightleftarrows a}}{{}^X \omega_{s \rightleftarrows a}}e^{-\beta q_R^{s \rightleftarrows a}} \\
&=\frac{{}^L \omega_{a \rightleftarrows s}}{{}^X \omega_{s \rightleftarrows a}}+\frac{{}^R \omega_{a \rightleftarrows s}}{{}^X \omega_{s \rightleftarrows a}}
=\frac{{}^X \omega_{a \rightleftarrows s}}{{}^X \omega_{s \rightleftarrows a}}
=e^{-\sigma_{FB}^{s \rightleftarrows a}},
\label{AD2}
\end{aligned}
\end{equation}
Eq. \eqref{AD1} becomes
\begin{equation}
\begin{aligned}
&\langle e^{-\beta Q_{FB}^r+\delta I_{FB}} \rangle =P^{m}_s\{T_{fb}^{s \to a}e^{-\sigma_{FB}^{s \to a}+\delta I_{FB}^{s \to a}}+T_{fb}^{s \to s}e^{\delta I_{FB}^{s \to s}}\}
+P^{m}_a\{T_{fb}^{a \to s}e^{-\sigma_{FB}^{a \to s}\delta I_{FB}^{a \to s}}+T_{fb}^{a \to a}e^{\delta I_{FB}^{a \to a}}\} \\
&=\langle e^{-\sigma_{FB}+\delta I_{FB}} \rangle =
P(\tau_{fb})_aT_{fb}^{a \to s}+P(\tau_{fb})_sT_{fb}^{s \to s}+P(\tau_{fb})_sT_{fb}^{s \to a}+P(\tau_{fb})_a T_{fb}^{a \to a}=P(\tau_{fb})_a+P(\tau_{fb})_s=1,
\label{AD3}
\end{aligned}
\end{equation}
where on the first line we used $\delta I_{FB}^{s \to a}=\ln P(\tau_{fb})_a / P^{m}_s$ and $T_{fb}^{s \to a}/T_{fb}^{a \to s}={}^X \omega_{fb}^{s \to a}/{}^X \omega_{fb}^{a \to s}=e^{\sigma_{FB}^{s \to a}}$ and on the second line $T_{fb}^{s/a \to a/s}=1-T_{fb}^{s/a \to s/a}$.

Furthermore, using Eq. \eqref{AD2} we obtain:
\begin{equation}
\begin{aligned}
\langle e^{-\beta Q_{FB}^r+\sigma_{FB}} \rangle & =P^{m}_s\{T_{fb}^{s \to a}[P_{L|s \to a}e^{-\beta q_L^{s \to a}}+P_{R|s \to a}e^{-\beta q_R^{s \to a}}]e^{\sigma_{FB}^{s \to a}} + T_{fb}^{s \to s}e^{0}\}\\
&+P^{m}_a\{T_{fb}^{a \to s}[P_{L|a \to s}e^{-\beta q_L^{a \to s}}+P_{R|a \to s}e^{-\beta q_R^{a \to s}}]e^{\sigma_{FB}^{a \to s}}+ T_{fb}^{a \to a}e^{0}\}\\
&=P^{m}_s[T_{fb}^{s \to a}+T_{fb}^{s \to s}]e^0+P^{m}_a[T_{fb}^{a \to s}+ T_{fb}^{a \to a}]e^0=P^{m}_s+P^{m}_a=1.
\label{AD4}
\end{aligned}
\end{equation}
Thus by combining Eqs. \eqref{AD3} and \eqref{AD4} we obtain
\begin{equation}
\begin{aligned}
\langle e^{-\beta Q_{FB}^r+\delta I_{FB}} \rangle
=\langle e^{-\sigma_{FB}+\delta I_{FB}} \rangle
=\langle e^{-\beta Q_{FB}^r+\sigma_{FB}} \rangle 
=1.
\end{aligned}
\end{equation}

\section*{APPENDIX D: DERIVATION OF EQ. \eqref{eq:IFTMeas}}

The integral fluctuation theorem for the measurement entropy is given by
\begin{equation}
\begin{aligned}
&\langle e^{-\beta Q_M+\delta I_M} \rangle =P^{i}_s[T_{m}^{s \to a}e^{-\beta q_Y^{s \to a}+\delta I_M^{s \to a}}+T_{m}^{s \to s}e^{\delta I_M^{s \to s}}]
+P^{i}_a[T_{m}^{a \to s}e^{-\beta q_Y^{a \to s}+\delta I_M^{a \to s}}+T_{m}^{a \to a}e^{\delta I_M^{a \to a}}] \\
&=P^{m}_a T_m^{a \to s}+P^{m}_sT_m^{s \to s}+P^{m}_sT_m^{s \to a}+P^{m}_aT_m^{a \to a}=P^{m}_s+P^{m}_a=1,
\end{aligned}
\end{equation}
where we used the local detailed balance condition for the transition probabilities, $T_m^{s \to a}/T_m^{a \to s}={}^Y \omega_m^{s \to a}/{}^Y \omega_m^{a \to s}=e^{\beta q_Y^{s \to a}}$, $\delta I_M^{s \to a}=\ln P^{m}_a / P^{i}_s$ and the conservation of probability. 

\end{widetext}

\bibliography{seb_adlib}
\end{document}